\documentclass[12pt,letterpaper]{JHEP}
\usepackage{amsmath,amssymb}
\input epsf.tex

\newcommand\IR{\relax{\rm I\kern-.18em R}}
\newcommand\ee{\end{equation}}
\newcommand\slr{$SL(2,\IR)$}
\newcommand\be[1]{\begin{equation} \label{#1}}

\newcommand\skipthis[1]{{}}

\newcommand\ct[1]{{``{#1},''}}
\newcommand\bt[1]{{\em {#1},}}

\newcommand\p{\ensuremath{\partial}}
\newcommand\vx{\ensuremath{\Vec{x}}}
\newcommand\vu{\ensuremath{\Vec{U}}}
\newcommand\vw{\ensuremath{\Vec{W}}}
\newcommand\vq{\ensuremath{\Vec{R}}}
\newcommand\vp{\ensuremath{\Vec{P}}}
\newcommand\vv{\ensuremath{\Vec{v}}}
\newcommand\vn{\ensuremath{\Vec{\p}}}
\newcommand\norm[1]{\ensuremath{\lvert{#1}\rvert}}
\newcommand\nxa[1]{\ensuremath{\norm{\vx-(\vx_{#1}+\vv_{#1}t)}}}
\newcommand\spsi[1]{\ensuremath{\frac{Q_{#1}}{\nxa{A}^2}}}

\title{Superconformal Multi-Black Hole Quantum Mechanics}
\author{Jeremy Michelson\thanks{After Sept.\ 1: New High
       Energy Theory Center; Rutgers University;\skipthis{126 Frelinghuysen Road;}
       Piscataway, NJ \ 08854.}\\
        Department of Physics \\ University of California \\ 
            Santa Barbara, CA \ 93106  \\ {\rm $\qquad$ and} \\
        Jefferson Physical Laboratories \\ Harvard University \\
            Cambridge, MA \ 02138}
\author{Andrew Strominger \\
        Jefferson Physical Laboratories \\ Harvard University \\
            Cambridge, MA \ 02138}
\abstract{The quantum mechanics of $N$ slowly-moving 
charged BPS black holes in 
five-dimensional ${\cal N}=1$ supergravity is considered. 
The moduli space metric of the $N$ black holes is derived and shown 
to admit $4$ supersymmetries. A near-horizon limit is found 
in which the dynamics of widely separated black holes decouples from 
that of strongly-interacting, near-coincident black holes. 
This decoupling suggests that the quantum states supported in the
near-horizon moduli space
can be interpreted as  internal states of a single composite black hole 
carrying all of the charge. 
The near-horizon theory is shown to have an enhanced $D(2,1;0)$ 
superconformal symmetry. 
Eigenstates of the Hamiltonian $H$ of the near-horizon theory are 
ill-defined due to noncompact regions of the
moduli space corresponding to highly redshifted near-coincident black
holes. It is argued that one should consider, instead of $H$ 
eigenstates, eigenstates of $2L_0= H+K$, where $K$ is the generator of 
special conformal transformations. The result is a well defined Hilbert
space with a discrete spectrum describing the $N$-black hole dynamics.} 

\preprint{hep-th/9908044 \\ HUTP-99/A047}

\begin{document}

\section{Introduction} \label{sec:intro}

One of the simplest examples of conformally-invariant quantum 
mechanics \cite{af} is given by the single-particle Hamiltonian
\be{eq:kds}
H= \frac {p^2}{2}+\frac{g}{2 x^2}.
\end{equation}
This system has a generator of dilations
\be{eq:krs}
D= \tfrac{1}{2} (px+xp)
\end{equation}
and special conformal transformations
\be{eq:krds}
K=\frac{ x^2}{2}.
\end{equation}
These operators obey the \slr\ conformal algebra
\begin{align} \label{eq:krdss}
[D,H]&=2iH,&[D,K]&=-2iK,&[H,K]&=-iD.
\end{align}
Of course since $D$ and $K$ do not commute with the Hamiltonian, 
they are not symmetries of the theory in the usual sense. Rather 
\slr\ can be used to relate states of different energies. 

In fact the theory defined by $H$ is hard to make sense of 
because there is no ground state. The wave function wants to spread out to
infinite $x$. Many years ago \cite{af} 
de Alfaro, Fubini and Furlan (DFF) 
suggested that one should consider, instead of $H$ eigenstates, eigenstates
of 
\be{eq:kos}
L_0=\frac{1}{2}(H+K)=\frac{ p^2}{4}+\frac{g}{4 x^2}+\frac{x^2}{4}.
\end{equation}
$L_0$ has a well behaved 
discrete spectrum of normalizable 
eigenstates.\footnote{The spectrum is
generated with $2L_{\pm 1}=H-K\mp iD$, which obey $[L_1,L_{-1}]=2L_0$ and 
$[L_0,L_{\pm 1}]=\mp L_{\pm 1}$.} 

This same problem reappeared recently in a 
new guise \cite{bhaf,azc,gt,kall}. 
Consider a charged BPS particle in a stationary trajectory 
a fixed but small distance from the horizon of an extremal Reissner-Nordstr\"{o}m 
black hole. The near-horizon  geometry of extremal Reissner-Nordstr\"{o}m 
is $AdS_2\times S^2$, so in this region the particle follows 
a trajectory of constant
$r$ in the $AdS_2$ coordinates
\be{eq:rds}
ds^2=-r^2dt^2+\frac{dr^2}{r^2}.
\end{equation}
Non-relativistic radial motion of this particle is described by the 
Hamiltonian $H$ of equation \eqref{eq:kds} with $g=0$. $g$ 
becomes nonzero when $S^2$ angular momentum is added. 
In this context the \slr\ symmetry generated by $H$, $D$ and $K$ 
is nothing but the \slr\ isometries of $AdS_2$.  

\FIGURE[t]{\leavevmode\epsfxsize=.66\hsize\epsfbox{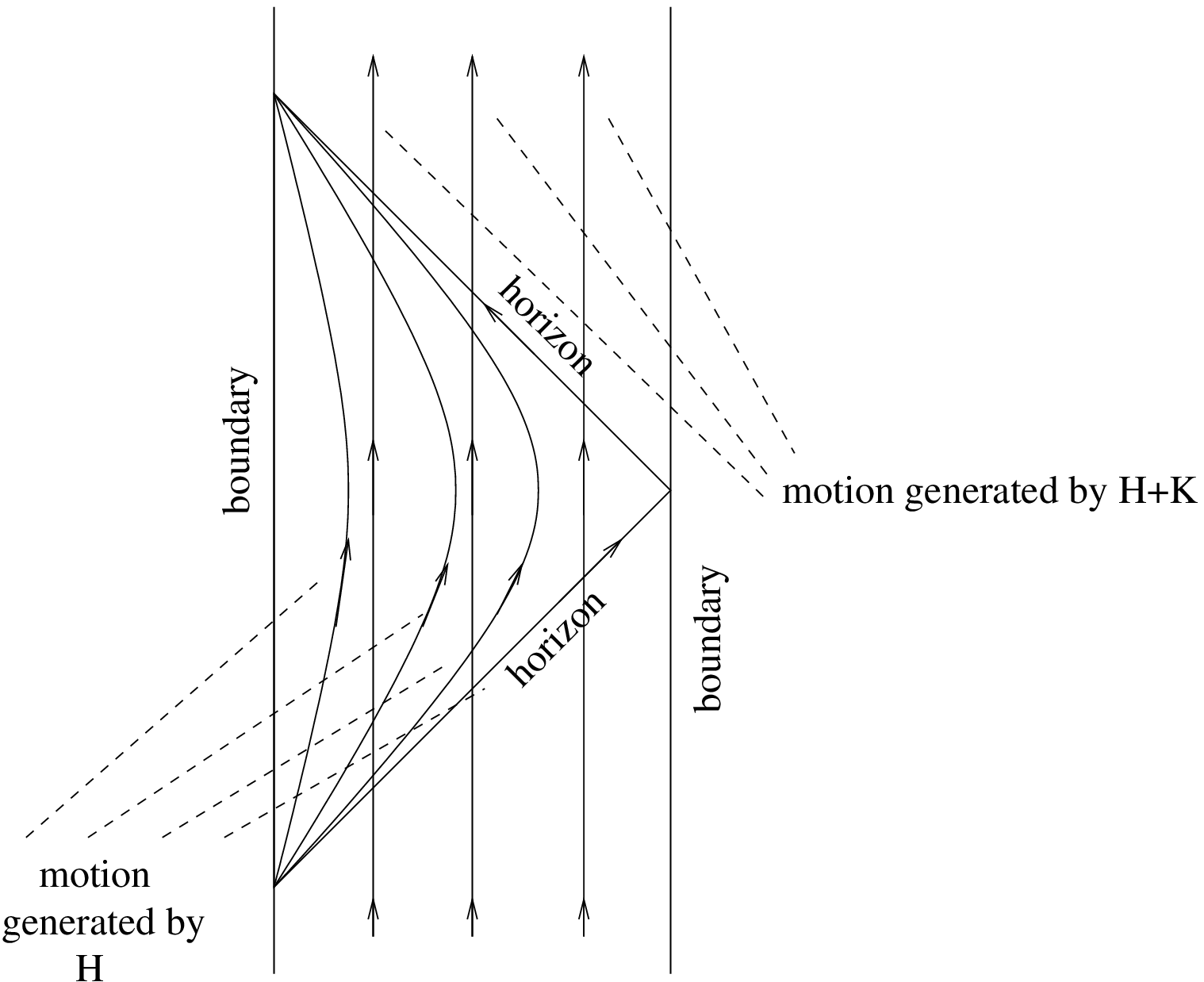}
\caption{The $AdS_2$ time coordinate conjugate to $H$ is badly behaved at the
horizon, but the time conjugate to $H+K$ is a good global
coordinate.}\label{fig:hvsl0}}

The absence of a good ground state for $H$ in this context is the 
familiar statement that there is a divergent 
continuum of very low-energy states
corresponding to highly redshifted trajectories very near the horizon
at $r \to 0$.
The origin of this problem is that the time conjugate to $H$ is not 
a good global time coordinate on $AdS_2$. On the other hand 
$L_0=\frac{1}{2}(H+K)$ is conjugate to a good global time coordinate. 
So, as pointed out in \cite{bhaf} and illustrated in
figure~\ref{fig:hvsl0}, in the black hole context the
DFF suggestion simply amounts to an improved choice of 
time coordinate. 

In this paper we apply this philosophy to the problem of the quantum
mechanics of $N$ slowly-moving five-dimensional
BPS black holes, or equivalently quantum
mechanics on the $N$-black hole moduli space.%
\footnote{The $N=2$ case in four dimensions was analyzed in~\cite{trasch}.}
This moduli space has two 
types of noncompact regions into which the wave function spreads,
preventing the existence of a normalizable ground state. The first 
such region is asymptotically flat ($\mathbb{R}^{4N}$ in five dimensions) 
corresponding to widely 
separated black holes. We show that in a low-energy, near-horizon limit,
this noncompact region decouples from the near-horizon region 
of strongly interacting black holes.  The decoupling of this near-horizon
region suggests that quantum states supported in the near-horizon region
can be interpreted as internal states of a single composite black hole, 
and that the number of such states is related to the black hole entropy. 
However the quantum Hilbert space is again hard to make sense of 
because of
noncompact regions of the moduli space corresponding to {\em near-coincident}
black holes. The continuum of low-energy states again arise because 
of large redshifts for near-coincident black holes. 
For the two-black hole case ($N=2$) the problem is equivalent
to that of a BPS particle near the black hole horizon, and the noncompact
region of the moduli space is exactly the troublesome $r \to 0$ 
region discussed above. 

This noncompact region can be eliminated by a generalization of the 
DFF suggestion. We show that the near-horizon moduli space 
has an \slr\ (more precisely $D(2,1;0)$) symmetry. Trading $H$ for $H+K$, 
the special conformal generator $K$ 
acts as a potential on the moduli space which diverges in all 
the noncompact directions, preventing the indefinite 
spread of the quantum wave functions. 
We note that unlike the $2$-body case, trading $H$ for $H+K$
cannot (as far as we know) be viewed as a new choice of time coordinate, and 
hence the $N$-body generalization is rather nontrivial. 
Nevertheless the result of the trade is a normalizable ground state
and a well defined  $4N$-body quantum mechanics.

It is worth emphasizing that no general argument that we know of 
guaranteed the existence of an \slr\ symmetry in the near-horizon limit. 
The existence of a dilational symmetry at low energies 
follows simply from dimensional arguments.  The 
existence of a special conformal generator $K$ however requires 
a very particular form of the geometry.

The theory we specifically consider is five-dimensional $N=1$ (8 supercharges)
supergravity with no vector multiplets. This can be obtained 
by compactification of $M$-theory on a Calabi-Yau with 
$b_2=1$. The black holes carry graviphoton charge. They can be described as 
M2-branes wrapping the Calabi-Yau 2-cycle. Other types of black holes---for
example those appearing in theories with more supercharges or vector
multiplets---have a different structure typically with 
extra degrees of freedom.  We do not know if the \slr\ symmetry exists in
all cases.

The outline of this paper is as follows. In section~\ref{sec:bhs} we
give a detailed derivation of the moduli space metric for 
$N$ BPS black holes in five dimensions, completing previous work 
on the subject \cite{shir,dkjm,gk,gps} and in particular exhibiting the 
requisite four supersymmetries.  The problem of finding the supersymmetries 
in four dimensions remains unsolved (although see~\cite{gps}). In
section~\ref{sec:nh} we find the near-horizon limit of the moduli space geometry. 
Using results of~\cite{jmas}, the $D(2,1;0)$ conformal symmetry of the
near-horizon moduli space
quantum mechanics is displayed in section~\ref{sec:su}.  Concluding comments
on the possible role of this near-horizon 
quantum mechanics as a dual description of black holes, the relation 
to M-theory constructions and the $AdS/CFT$ correspondence~\cite{jm}, 
are in section~\ref{sec:conc}.

\section{Derivation of the $D=5$ Multi-Black Hole Moduli Space} \label{sec:bhs}

In this section we derive the 
moduli space of $N$ BPS black holes in five-dimensional $\mathcal{N}=1$ 
supergravity 
with a single $U(1)$ charge, coupled to the graviphoton, and no 
vector multiplets. Neutral hypermultiplets, if present, do not affect the
discussion because they decouple. The relevant part of the 
action is~\cite{cremmer}  
\begin{equation} \label{eq:puresugra}
S = \frac{M_p}{2} \int d^5x \sqrt{g} \bigl[M_p^2R - \frac{3}{4} F^2]
 + \frac{\lambda}{2} \int A \wedge F \wedge F,
\end{equation}
where $\lambda = 1$ (but we keep it general for most of the
discussion) and
we have normalized the vector field in an unconventional but convenient
way.

The static, $N$ extremal black hole solutions are given by
\begin{subequations} \label{eq:bhsoln}
\begin{gather}
\label{eq:bhsolnds}
ds^2 = -\psi^{-2} dt^2 + \psi d\vx^2, \\
\intertext{and}
\label{eq:bhsolnA}
A = M_p \psi^{-1} dt, \\
\intertext{where $\psi$ is the harmonic function on $\mathbb{R}^4$}
\label{eq:bhsolnpsi}
\psi = 1 + \sum_{A=1}^N \frac{Q_AL_p^2}{\norm{\vx-\vx_A}^2},
\end{gather}
\end{subequations}
$L_p=\frac{1}{M_p}$ and $\vx_A$ is the $\mathbb{R}^4$ coordinate of the
$A$th black hole with charge
$Q_A$.

Supersymmetry%
\footnote{Actually, this is more general than supersymmetry, since it holds for
any value of $\lambda$ in the action.}\ 
implies a no-force condition between the black holes, and thus the $\vx_A$
are moduli for the set of states described by equations~\eqref{eq:bhsoln}.
The effective action governing slowly-moving black holes is therefore
described by the metric on this moduli space; we will derive this metric
following~\cite{fe,shir,dkjm}, and show that (for $\lambda=1$) it has the
properties expected for $\mathcal{N}=4B$ supersymmetric quantum
mechanics~\cite{gps,jmas}.

The first step in the derivation of the effective action,
following \cite{fe} is to add the
source terms
\begin{equation} \label{eq:source}
S_{\text{source}} = -\frac{6\pi^2}{L_p} \sum_{A=1}^N Q_A \int ds_A 
+ 6\pi^2 \sum_{A=1}^N Q_A \int  A_\mu dx_A^\mu,
\end{equation}
to the action, where $ds_A$ is the line element of the center of the $A$th
black hole.  In the remainder of this section we set $L_p=1$. 
The terms in equation~\eqref{eq:source} are necessary because the equations
of motion
following from \eqref{eq:puresugra} are not satisfied at the timelike
singularities
located behind the $N$ black hole event
horizons at $\vx=\vx_A$.  These source terms will affect the
calculation as we have set it up, because we work in a gauge in which the zero
modes corresponding to motion of the $A$th black hole do not vanish 
at $\vx=\vx_A$. In principle it should be possible to choose a gauge 
in which the zero mode vanishes at $\vx=\vx_A$, 
and hence avoid adding the terms 
\eqref{eq:source}. However we have found the integrals most tractable in the
chosen gauge.  In \cite{fe} the moduli space metric (in the
four-dimensional Einstein-Maxwell theory) was computed by first
smoothing out these sources and then letting them approach delta functions
at the end of the calculation. We shall not find this cumbersome 
procedure necessary.     

An ansatz describing the perturbation of the black hole
solution~\eqref{eq:bhsoln} by non-zero, but small velocities 
to linear order is 
\begin{subequations} \label{eq:pertbh}
\begin{gather}
\label{eq:pertbhds}
ds^2 = -\psi^{-2} dt^2 + \psi d\vx^2 + 2 \psi^{-2} \vq \cdot d\vx dt, \\
\intertext{and}
\label{eq:pertbhA}
A = \psi^{-1} dt + (\vp - \psi^{-1} \vq) \cdot d\vx,
\end{gather}
\end{subequations}
where $\vp$ and $\vq$ are quantities that are first order in velocities.
Also, in equation~\eqref{eq:bhsolnpsi}, $\vx_A$ is 
replaced with $\vx_A+\vv_At$.
Note that this is the most general galilean-invariant ansatz to linear
order.

Inserting the ansatz~\eqref{eq:pertbh} into
equation~\eqref{eq:source} gives, keeping only terms up to second order in
velocities,
\begin{equation} \label{eq:worksource}
\begin{split}
S_{\text{source}} &= 
-{6\pi^2} \sum_{A=1}^N Q_A\!\!\int\!dt \, \psi^{-1} \Bigl(1 -
\vq \cdot
\vv_A - \frac{1}{2} \psi^3 \vv_A^2 \Bigr)
\\ & \qquad \qquad
+ 6\pi^2 \sum_{A=1}^N Q_A\!\int\!dt \Bigl(\psi^{-1} + \vp \cdot \vv_A -
\psi^{-1} \vq \cdot \vv_A \Bigr)
\\ & = \frac{1}{2} \sum_{A=1}^N Q_A \int dt \Bigl( 6 \pi^2 \psi^2 \vv_A^2 +
12 \pi^2
\vv_A \cdot \vp \Bigr).
\end{split}
\end{equation}
The total action,~\eqref{eq:puresugra}
and~\eqref{eq:source} is, to second order in
velocities,
\begin{multline} \label{eq:pact}
S = \frac{1}{2} \int d^5x \Bigl\{ 
6 \pi^2 \psi^2 \sum_{A=1}^N Q_A \delta^{(4)}(\vx - (\vx_A+\vv_A t))\vv_A^2 
+ 12 \pi^2 \sum_{A=1}^N Q_A \delta^{(4)}(\vx - (\vx_A+\vv_A t)) \vv_A \cdot \vp
\Bigr. \\ \Bigl.
+ 3 \p_t \vp \cdot \vn \psi 
- \frac{3}{4} \psi^{-1} (\p_i P_j - \p_j P_i)^2
+ 3 \psi^{-2} (\p_i P_j - \p_j P_i) \p_i R_j
- \frac{1}{2} \psi^{-3} (\p_i R_j - \p_j R_i)^2
\Bigr. \\ \Bigl.
- 3 \psi (\p_t \psi)^2
- 3 \lambda \psi^{-1} \epsilon^{ijkl} \p_i P_k \p_j P_l
+ 3 \lambda \psi^{-2} \epsilon^{ijkl} \p_i P_k \p_j R_l
- \lambda \psi^{-3} \epsilon^{ijkl} \p_i R_k \p_j R_l
+ \text{t.d.}
\Bigr\},
\end{multline}
where $i,j=1,\dots,4$ are spatial indices and t.d. stands for total 
derivative.

Our immediate goal is to eliminate $\vp$ and $\vq$ from the 
effective action \eqref{eq:pact} to get an action involving only 
$\vv_A$ as a dynamical variable.   
This requires the equations of motion which yield%
\skipthis{\footnote{These
solutions are independent of $\lambda$.  While
the Chern-Simons term in the action, proportional to
$\lambda$, is non-zero, it turns out to give only coexact contributions to
the equations for $d^\dagger dP$ and $d^\dagger dR$, and so are absorbed
into the functions of
integration.  (This is the analog of the statement in~\cite{dkjm}---which
considered only $\lambda=0$---that
equations~\eqref{eq:eomq} and~\eqref{eq:eomp} should, in principle, have arbitrary
functions of integration on the right-hand sides, but that demanding
$d^2P=0=d^2R$ results in homogeneous equations for the functions of
integration, so that they can be set to zero.)}}
\begin{subequations} \label{eq:eoms}
\begin{gather}
\label{eq:eompsi}
\vn^2 \psi = -4 \pi^2 \sum_{A=1}^{N} Q_A \delta^{(4)}
(\vx-\vx_A-\vv_At), \\
\label{eq:eomq}
dR = -3 \psi^2 \sum_{A=1}^N d\spsi{A} \wedge v_A, \\
\intertext{and}
\label{eq:eomp}
dP = -2 \psi \sum_{A=1}^N d\spsi{A} \wedge v_A.
\end{gather}
\end{subequations}
Equation~\eqref{eq:eompsi} follows from the $A_0$ (and $g_{00}$) equation
of motion and
is solved by equation~\eqref{eq:bhsolnpsi}.  Equations~\eqref{eq:eomq}
and~\eqref{eq:eomq}
follow from linear combinations of the $g_{0i}$ and $A_i$ 
equations of motion.

Note that $\vp$ and $\vq$ appear in the effective action \eqref{eq:pact}
only as $dP$ and $dR$, except in the
second and third terms.  In order to apply
equations~\eqref{eq:eoms}, we need to do the following
manipulation on
those terms:
\begin{equation}
\begin{split} \label{eq:workonp}
12 \pi^2 & \sum_{A=1}^N Q_A \delta^{(4)}(\vx - (\vx_A+\vv_A t)) \vv_A \cdot \vp
+ 3 \p_t \vp \cdot \vn \psi
\displaybreak[0] \\
&= -3 \vp \cdot \vn \p_t \psi 
- 3 \sum_{A=1}^N \vn^2 \frac{Q_A}{\nxa{A}^2}
\vv_A \cdot \vp
+ \text{t.d.} \displaybreak[0] \\
&= 3 \vp \cdot \vn \sum_{A=1}^N \vv_A \cdot \vn
       \frac{Q_A}{\nxa{A}^2}
-3 \sum_{A=1}^N \vn^2 \frac{Q_A}{\nxa{A}^2} \vv_A \cdot \vp
+ \text{t.d.} \displaybreak[0] \\
&= -3 \sum_{A=1}^N \vv_A \cdot \vn \vp \cdot \vn \spsi{A}
+ 3 \sum_{A=1}^N \vn \spsi{A} \cdot \vn \vp \cdot \vv_A 
+ \text{t.d.} \displaybreak[0] \\
&=- 3 \sum_{A=1}^N v_{Ai} (\p_i P_j - \p_j P_i) \p_{Aj} \psi
+ \text{t.d.}
\end{split}
\end{equation}
Now the only $P$-dependence of the action is in the form $dP$, and
we can apply equations~\eqref{eq:eoms}.  This yields
\begin{multline} \label{eq:actionpref}
S = \frac{1}{2}\int d^5x \Bigl\{
-\frac{3}{2} \psi^2 \sum_{A=1}^N \vn_A^2 \psi \vv_A^2
-3 \psi \sum_{A,B=1}^N \vv_A \cdot \vv_B \vn_A \psi \cdot \vn_B \psi
+3 \psi \sum_{A,B=1}^N \vv_A \cdot \vn_B \psi \vv_B \cdot \vn_A \psi \\
-3 \psi \sum_{A,B=1}^N \vv_A \cdot \vn_A \psi \vv_B \cdot \vn_B \psi
-3 \lambda \psi \epsilon^{ijkl} \sum_{A,B=1}^N \p_{Ai} \psi v_{Aj} \p_{Bk} 
\psi v_{Bl}
+ \text{t.d.} \Bigr\}
\end{multline}
or equivalently 
\begin{equation} \label{eq:actionwithf}
S = -\frac{1}{4} \int dt \sum_{A,B=1}^N
\bigl( \delta^{ij}\delta_{kl} + \delta^i_k
\delta^j_l - \delta^i_l \delta^j_k + \lambda \epsilon^{ij}{_{kl}} \bigr)
\p_{Ai} \p_{Bj} \Bigl( \int d^4x \psi^3 \Bigr) v^{Ak} v^{Bl}.
\end{equation}
Precisely in the supersymmetric case, when $\lambda=1$,%
\footnote{$\lambda=-1$ differs by a reflection, and so we also get a
supersymmetric moduli space with the antiself-dual complex structures.}\ 
this can be
rewritten as
\begin{equation} \label{eq:actionsusic}
S = \frac{1}{4} \int dtv^{Ak} v^{Bl}
\bigl( \delta^{i}_{k} \delta^{j}_{l} + \sum_{r=1}^3 I^r_{k}{^{i}}
I^r_{l}{^{j}} \bigr)
\p_{Ai} \p_{Bj} L,
\end{equation}
where 
\be{eq:ldef}
L=-\int d^4x \psi^3
\end{equation}
and $I^r$, $r=1,2,3$ are the natural triplet of self-dual
complex structures on $\mathbb{R}^4$.
In \cite{jmas} it was shown that any action of the form 
\eqref{eq:actionsusic} has 
(after appropriately  adding fermions) $\mathcal{N}=4B$ supersymmetry 
(which has an $SU(2)$ $R$-symmetry under which the bosons 
are complex doublets)  for 
any function $L$. Hence we conclude that the multi-black hole 
moduli space has  $\mathcal{N}=4B$ supersymmetry.

We note that the first term of equation~\eqref{eq:actionwithf} (i.e.\
the term with $\delta^{ij} \delta_{kl}$) has previously appeared in the
literature \cite{shir,dkjm} on the five-dimensional black hole moduli
space, but the last three terms (required by supersymmetry) 
were not previously included.

\section{The Near-Horizon Limit} \label{sec:nh}

In this section we describe the ``near-horizon limit'' 
of a collection of $N$ black holes, in which
\begin{equation} \label{eq:rs}
\frac{|\vx_A-\vx_B|^2}{L_p^2} \to 0,
\end{equation}
with charges of order one. 
This corresponds to taking the characteristic 
distances between the black holes 
to be much less than the Planck length. It also corresponds to 
kinetic energies much less than the Planck mass, because the 
relative kinetic energy of two black holes becomes highly redshifted when
they are near one another. 

\FIGURE[t]{\leavevmode\epsfxsize=1\hsize\epsfbox{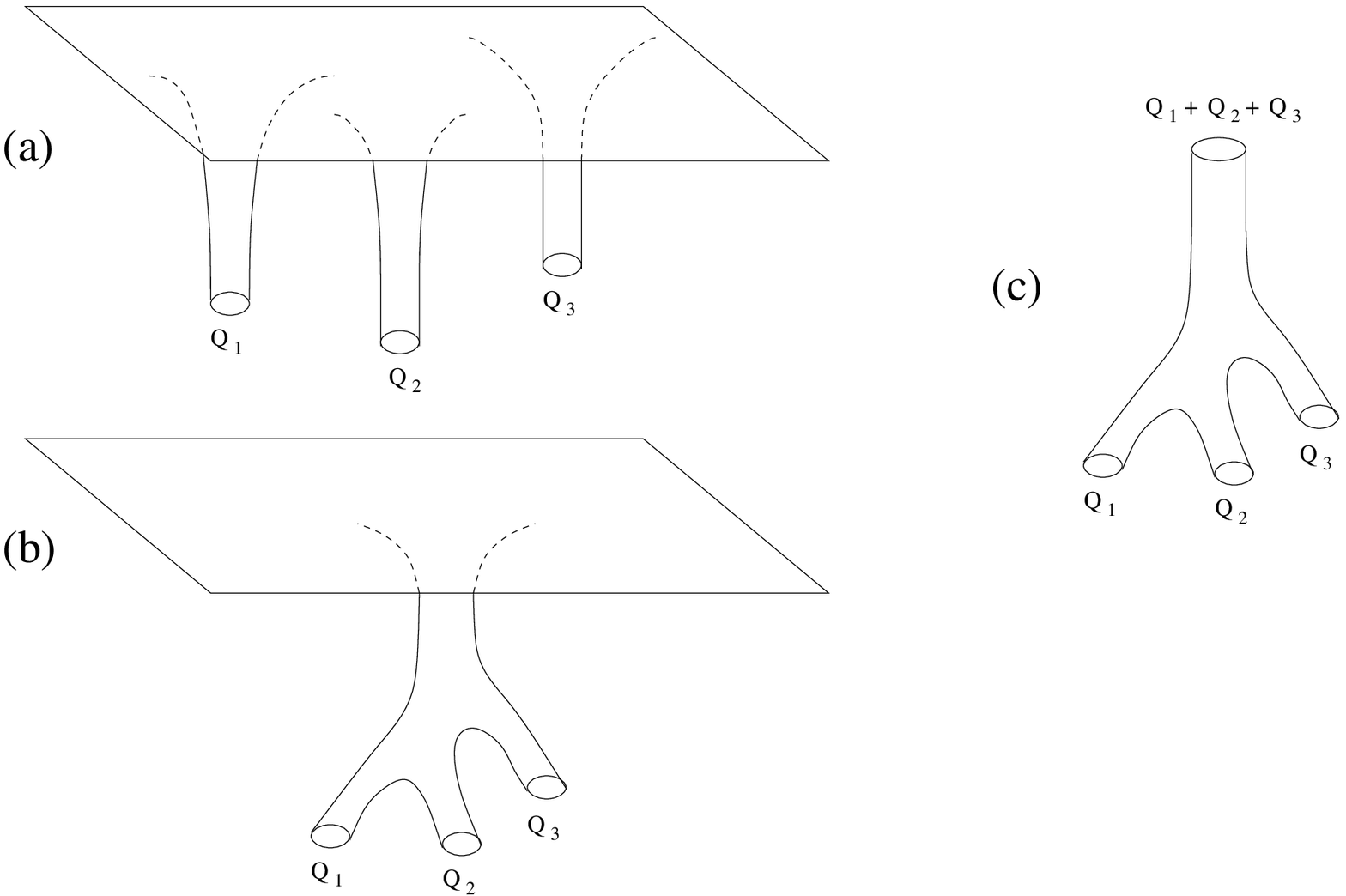}
\caption{(a) Widely separated black holes.  (b) Near-coincident black
holes.  (c) The near-horizon limit.}\label{fig:nhst}}

The limiting geometry can be derived by 
defining new coordinates
\begin{align} \label{eq:lki}
\vu_A&=\frac{\vx_A}{L_p^{3/2}},&\vu&=\frac{\vx}{L_p^{3/2}},
\end{align}
and then taking 
\be{eq:jrf}
L_p \to 0.
\end{equation}
The spacetime metric for the $N$-black hole solution then becomes
\be{eq:jpf}
 \frac{ds^2}{L_p^2} = -\psi^{-2} dt^2 + \psi d\vu^2, 
\end{equation}
with 
\be{eq:pgl}
\psi = \sum_{A=1}^N \frac{Q_A}{\norm{\vu-\vu_A}^2},
\end{equation}
An illustration of the resultant spatial geometry at a moment of fixed time
is given
in figure~\ref{fig:nhst} for $N=3$.  
Before the limit is taken (figure~\ref{fig:nhst}a), the geometry has an 
asymptotically flat region at large $\norm{\vx}$.  Near the limit 
(figure~\ref{fig:nhst}b), as the origin
is approached along a spatial trajectory, a single ``throat'' 
approximating that of a charge $\sum Q_A$ 
black hole is encountered. This throat region is 
an $AdS_2\times S^3$ geometry with radii of order $\sqrt {\sum Q_A}$.  
As one moves deeper inside the throat towards the horizon, the throat 
branches into smaller throats, each of which has smaller charge and 
correspondingly smaller radii. Eventually there are $N$ branches with
charge $Q_A$. At the end of each of these branches is an event horizon. 
When the limit is achieved (figure~\ref{fig:nhst}c), the asymptotically flat 
region moves off to
infinity. Only the charge $\sum Q_A$  ``trunk'' and the many branches remain. 

It is also of interest to consider the near-horizon limit of 
the moduli space geometry. This is simply\footnote{If the black holes are
identical, one must take a quotient with respect to the permutation group.} 
\be{eq:lml}
\frac{ds^2}{L_p^3}=\frac{1}{4}\bigl( \delta^{i}_{k} \delta^{j}_{l} + 
\sum_{r=1}^3 I^r_{k}{^{i}}
I^r_{l}{^{j}} \bigr)
\bigl(\p_{Ai} \p_{Bj} L\bigr) dU^{Ak} dU^{Bl},
\end{equation}
with 
\be{eq:ldf}
L=-\int d^4U \psi^3 
\end{equation}
and $\psi$ now given by \eqref{eq:pgl}. This is illustrated in 
figure~\ref{fig:nhms} for the case of $N=2$. Near the limit there is an 
asymptotically flat $\mathbb{R}^{4N}$ region corresponding to 
all $N$ black holes widely separated. This is connected 
to the near-horizon 
region where the black holes are strongly-interacting, by 
tube-like regions which become longer and thinner as the 
limit is approached. When the limit is achieved, the 
near-horizon region is severed from the tubes and the asymptotically 
flat region. The near-horizon metric \eqref{eq:lml} develops four zero 
eigenvectors $U^{Ak}=T^k$ corresponding to the decoupling of the four 
overall center of
mass coordinates.  For the case of two black holes the near-horizon region is 
just flat $\mathbb{R}^4$, with the tube touching at the origin. 

\FIGURE[t]{\leavevmode\epsfxsize=1\hsize\epsfbox{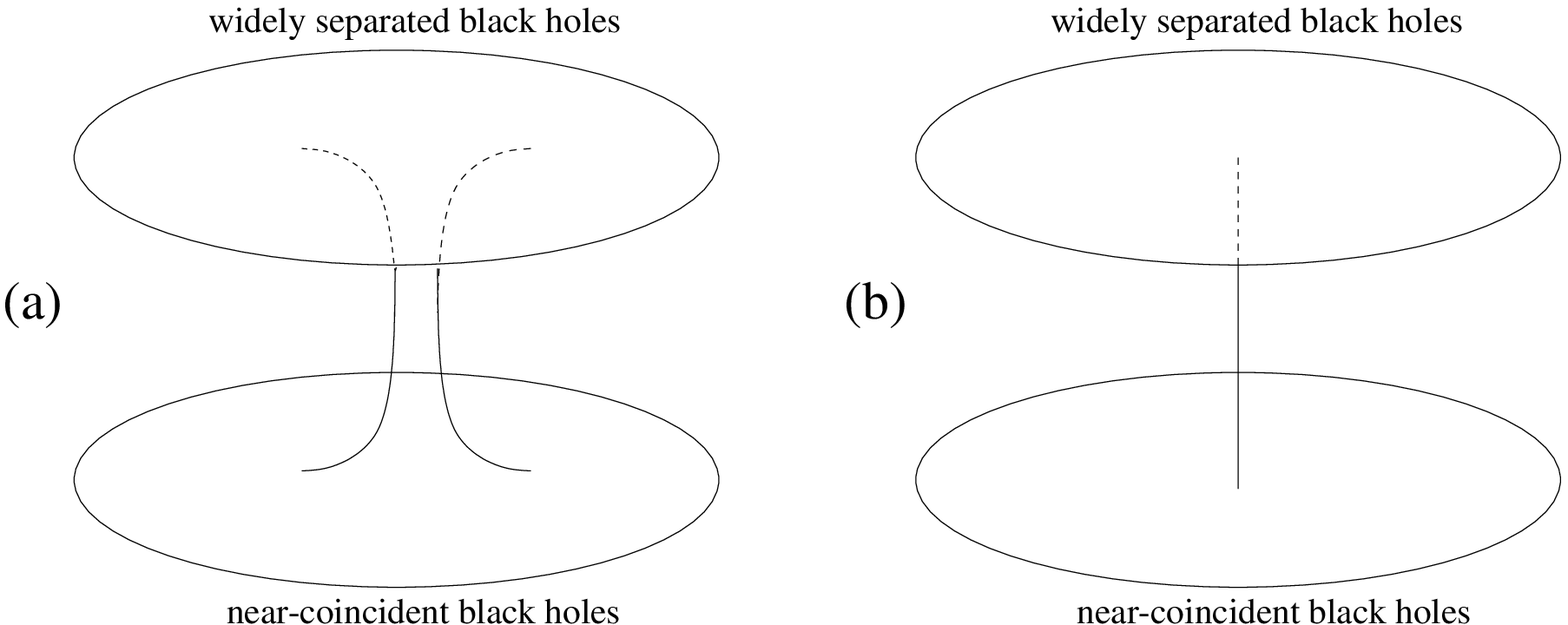}
\caption{(a) Regions of the two-black hole moduli space. (b) The
near-horizon limit.}\label{fig:nhms}}

\section{Conformal Symmetry} \label{sec:su}

The fact that the near-horizon moduli space geometry decouples 
from the rest of the moduli space,  suggests that the quantum states
on this moduli space may be interpreted as internal states of 
a single charge $\sum Q_A$ black
hole.  However an obstacle is immediately encountered in attempting to 
make sense of the quantum mechanics: the Hamiltonian has no ground state
because of the infinite volume ascribed to {\em near-coincident} black holes. 
Indeed, near $\vu_A=\vu_B$ the metric has a term 
\be{eq:gsi}
(Q_A^2Q_B+Q_B^2Q_A) \frac{\norm{d\vu_A-d\vu_B}^2 }{\norm{\vu_A-\vu_B}^4}.
\end{equation}
Defining $\vw_{AB}=\dfrac{\vu_A-\vu_B }{\norm{\vu_A-\vu_B}^2}$,
\eqref{eq:gsi} becomes 
\be{eq:gsi2}
 (Q_A^2Q_B+Q_B^2Q_A)\norm{d\vw_{AB}}^2.
\end{equation}
This is just the flat metric on $\mathbb{R}^4$, with infinity corresponding to the 
near-coincident limit (and the origin to infinite separation).  
Because of these infinite-volume regions, 
the quantum system has an infinite number of states below any finite 
excitation energy, and no normalizable ground state. 
This is in conflict with the 
usual stringy picture of a black hole as a system with a gap and 
a discrete spectrum. 

In this section we will show that the near-horizon 
quantum mechanics has an \slr\ conformal symmetry (in fact a
$D(2,1;0)$ superconformal symmetry) with 
generators of time
translations $H$, 
dilations $D$ and special conformal transformations $K$.  
The Hilbert space can then be sensibly defined as the eigenstates of $H+K$, 
rather than of $H$ itself.

To see this conformal symmetry 
we begin by dividing the potential $L$ appearing in the metric
\eqref{eq:lml} up into pieces which 
represent the 1-body, 2-body and 3-body interactions 
\begin{subequations} \label{jsall}
\be{eq:jsi}
L_1=-\sum_{A=1}^N\int d^4U \frac{Q_A^3}{\norm{\vu-\vu_A}^6},
\end{equation}
\be{eq:jsia}
L_2=-3\sum_{A \neq B}^N\int d^4U 
\frac{Q_A^2Q_B}{\norm{\vu-\vu_A}^4\norm{\vu-\vu_B}^2},
\end{equation}
\be{eq:jsib}
L_3=-\sum_{A\neq B\neq C}^N\int d^4U \frac{Q_AQ_BQ_C}{\norm{\vu-\vu_A}^2
\norm{\vu-\vu_B}^2\norm{\vu-\vu_C}^2},
\end{equation}
\end{subequations}
with $L=L_1+L_2+L_3$. It is a curious (and unexplained) fact 
that there are no 4-body or higher interactions.   
$L_1$ is a divergent $\vu_A$-independent constant. This divergence 
does not enter the metric which involves only derivatives of 
$L$. Hence $L_1$ can be ignored. $L_2$ has logarithmic 
divergences near $\vu=\vu_A$.
With a cutoff $\delta$ the divergences 
go like $\frac{\ln \delta}{\norm{\vu_A-\vu_B}^2}$. These terms also
disappear after 
differentiating $L$ to form the metric. The remaining term, which does
contribute to the metric, is 
\be{eq:jsia2}
L_2=-6\pi^2\sum_{A \neq B}^N 
\frac{Q_A^2Q_B\ln \norm{\vu_A-\vu_B}}{\norm{\vu_A-\vu_B}^2}.
\end{equation} 
$L_3$ is finite and can be defined without a regulator. It follows that 
(unlike $L_2$) it
is a homogeneous polynomial of degree $-2$ obeying
\be{eq:lsf}
U^{Ai}\p_{Ai}L_3=-2L_3.
\end{equation} 
We further note that by rotational invariance
\be{eq:lgd}
U^{Ai}I_i^{rj}\p_{Aj}L_2=0=U^{Ai}I_i^{rj}\p_{Aj}L_3.
\end{equation} 

First we look for a dilational symmetry of the 
2-body and 3-body metrics $g_2$ and $g_3$. 
It is easy to see that the vector field 
\be{eq:ji}
D^{Ai}\p_{Ai}=-U^{Ai}\p_{Ai}
\end{equation}
generates a dilational symmetry 
under which the metric transforms as ${\cal L}_Dg_{ab}=2g_{ab}$. 
Although $L_2$ (equation~\eqref{eq:jsia2}) is not
homogeneous because of the logarithm, the anomalous piece does not 
contribute in the metric.

It was shown in \cite{jmas} that 
in order to have a special conformal symmetry the one-form 
\be{eq:opi}
 D_{Ai}dU^{Ai}
\end{equation}
must be closed. 
The one-form is obtained from the vector field by 
lowering the index with the metric $g=g_2+g_3$. A little algebra 
using equation~\eqref{eq:lml} reveals that any metric constructed from a potential 
obeying the weight $-2$ scaling relation \eqref{eq:lsf} has 
$U^{Ai}$ as a zero eigenvector \cite{jmas}. Therefore
\be{eq:ldo}
D_{3Ai}=-g_{3AiBj}U^{Bj}=0.
\end{equation}
Explicit computation then gives (suppressing factors of $L_p$)
\be{eq:ldyo} 
D_{Ai}dU^{Ai}=-g_{2AiBj}U^{Bj}dU^{Ai}=d\left[6 \pi^2\sum_{A \neq B}^N 
\frac{Q_A^2Q_B}{\norm{\vu_A-\vu_B}^2}\right].
\end{equation}
Hence $D_{Ai}dU^{Ai}$ is closed, and as shown in \cite{jmas},
the quantity in square brackets is the generator of special conformal
transformations
\be{eq:lyo} 
K=6 \pi^2\sum_{A \neq B}^N 
\frac{Q_A^2Q_B}{\norm{\vu_A-\vu_B}^2}.
\end{equation}
This geometry meets all the criteria described in \cite{jmas} 
for the full $D(2,1;0)$ superconformal symmetry. This group is 
the special case of the $D(2,1;\alpha)$ superconformal groups 
for which there is an $SU(1,1|2)$ 
subgroup.%
\footnote{Specifically, $D(2,1;0)$ is the semi-direct sum of
$SU(1,1|2)$ and $SU(2)$.}
The appearance of $SU(1,1|2)$, in the near-horizon limit of the spacetime
geometry of a single black hole, was observed
in~\cite{gmt}.

It is worth noting that while the existence of the dilational symmetry was 
more or less guaranteed by the nature of the near-horizon limit, 
the same cannot be said for the special conformal symmetry. Indeed if the 
coefficients of the four terms in \eqref{eq:actionwithf} are 
perturbed, the dilational symmetry remains 
but the special conformal symmetry generically disappears.

\section{Discussion} \label{sec:conc}

We have given a manifestly supersymmetric presentation 
of quantum mechanics on the five-dimensional $N $-black hole moduli 
space. At low energies, the dynamics of near-coincident black holes
decouples from that of widely separated black holes, and an enhanced
\slr\ symmetry appears. There are 
noncompact regions of this near-horizon 
moduli space corresponding to 
coincident black holes. These regions are eliminated by the 
potential $K$ in the modified Hamiltonian $L_0=\frac{1}{2}(H+K)$,
which is singular at the boundary of the noncompact regions. 
We then have an apparently well defined, albeit complicated,  
quantum mechanics describing black holes which are 
neither widely separated nor coincident. 
A detailed description of the quantum states of this system 
remains to be found. 
For now we content ourselves
with a few comments on their possible interpretation.   

Let us return to the case of $M$-theory on Calabi-Yau 
with $b_2=1$. An alternate picture of the charge $N$ black hole is 
then an $M2$-brane holomorphically wrapped $N$ times around the 
K\"{a}hler class. In the quantum theory one has a quantum mechanics on 
the moduli space of $N$-wrapped holomorphic cycles. 
The black hole entropy should be given by the 
logarithm of the number of ground states of the $M2$-brane, 
which is roughly speaking the Euler character of the moduli space. 
While $AdS_2$/$ CFT_1$ duality
\cite{jm,asads,mms,ny,cm,gt,pkt,sprad,blum}
has eluded a proper understanding,
this quantum mechanics on the $M2$-brane moduli space should more or less be
the theory which lives on the boundary (boundaries?) of $AdS_2$.  
 
This moduli space has several components. In one component---which we will
refer to as the ``Higgs'' branch in an abuse of language
(since there is no gauge theory on the $M2$-brane)---one has a single
connected $M2$-brane
mapped holomorphically into the Calabi-Yau. There is one center of 
mass degree of freedom corresponding to transverse motion on the spatial 
$\mathbb{R}^4$. The quantum states on the analog of this branch for 3-branes on 
$K^3\times S^1$ and  5-branes on Calabi-Yau$\times S^1$ were successfully
related to black hole entropy in \cite{ascv} and \cite{msw}.\footnote{Note
however that these computations of the entropy 
required only the dimension of the moduli 
space, while membranes on a Calabi-Yau require the cohomology of the moduli 
space. This is much more difficult to compute, and to date there is 
no microscopic computation of the entropy of these black holes.}  In addition
there are ``Coulomb'' branches corresponding to disconnected
membranes. These branches have up to $N$ center-of-mass coordinates. 
Since we have only a single type of charge there is no way to lift these 
branches. The importance of these branches for understanding $AdS_2$ 
black holes was discussed in \cite{mms}. 

Na\"{\i}vely what has been described in this paper 
has $N$ center-of-mass coordinates and is therefore the Coulomb
branch. In analogy with the calculations in \cite{ascv,msw}
one might expect that the black hole entropy counts states on the 
Higgs branch, and so should not be captured by the Coulomb branch
considerations of this paper. However we would like to suggest 
an alternate interpretation. 
The Coulomb branch described herein divides into the far-Coulomb 
branch (containing widely separated black holes) and the near-Coulomb 
branch (containing near-coincident black holes). As we have seen, 
these two regions 
decouple completely 
at low energies. A natural conjecture is that the near-Coulomb 
branch is in fact a dual description of the Higgs branch, described above 
as the quantum mechanics of a single $N$-wrapped membrane. A 
related phenomenon occurs in the $D1$-$D5$ black hole, for which 
the near-Coulomb branch is a dual description of the 
small-instanton region of the Higgs branch---for a recent discussion,
see~\cite{bv}. 
If this is the case 
one does expect quantum mechanics in the geometry \eqref{eq:lml} 
to capture the black hole entropy.

\acknowledgments

 We are grateful to R.\ Britto-Pacumio, J.\ Maldacena, M.\ Spradlin, P.\
Townsend  and especially
G. Papadapolous for useful conversations. This work was supported in part
by an NSERC PGS B Scholarship and DOE grant DE-FG02-91ER40654.

\end{document}